\documentclass[aps,prl,preprint,superscriptaddress,showpacs,showkeys,]{revtex4}

\usepackage{graphicx}
\usepackage{dcolumn}
\usepackage{bm}
\usepackage{amsfonts}
\usepackage{amsmath}
\usepackage{amssymb}
\usepackage{amstext}
\usepackage{latexsym}
\usepackage{textcomp}
\usepackage{color}
\usepackage{tikz}
\makeatletter

\PassOptionsToPackage{normalem}{ulem}
\usepackage{ulem}

\providecolor{added}{rgb}{0,0,1}
\providecolor{deleted}{rgb}{1,0,0}

\makeatother

\begin{document}
\def\myfrac#1#2{\frac{\displaystyle #1}{\displaystyle #2}}


  	\newcommand{\fracd}[2]{%
		\frac{\displaystyle #1}{\displaystyle #2}}
	\newcommand\units[1]{%
  		\mathrm{~#1}}
	\newcommand\uni[1]{%
  		\mathrm{#1}}
	\newcommand{\modu}[1]{%
		\left|#1\right|}
	\newcommand{\E}[1]{%
		\times 10^{#1}}
	\newcommand\ee[1]{%
		\mathrm{e}^{#1}}
		

	\newlength\WidthLastColumn
	\setlength{\WidthLastColumn}{4.3cm}

%
%

\title{Role of the temperature instabilities for formation of nano-patterns upon single femtosecond laser pulses on gold}
\author{Evgeny L. Gurevich}
\email[Corresponding author: ]{gurevich@lat.rub.de}
\affiliation{Ruhr-Universit\"at Bochum, Chair of Applied Laser Technology,
Universit\"atsstra\ss e~150, 44801 Bochum, Germany}
\author{Yoann Levy}
\email[Corresponding author: ]{levy@fzu.cz}
\affiliation{HiLASE Centre, Institute of Physics AS CR, Za Radnic\'i 828, 25241 Doln\'i B\v{r}e\v{z}any, Czech Republic}
\author{Svetlana V. Gurevich}
\affiliation{Institut f\"ur Theoretische Physik
Westf\"alische Wilhelms-Universit\"at M\"unster
Wilhelm-Klemm-Stra\ss{}e 9
D-48149 M\"unster, Germany}
\author{Nadezhda M. Bulgakova}
\affiliation{HiLASE Centre, Institute of Physics AS CR, Za Radnic\'i 828, 25241 Doln\'i B\v{r}e\v{z}any, Czech Republic; S.S. Kutateladze Institute of Thermophysics SB RAS, 1 Lavrentyev ave., 630090 Novosibirsk, Russia}

\date{\today}

\begin{abstract}

In this paper we investigate whether the periodic structures on metal surfaces exposed to single ultrashort laser pulses can appear due to an instability induced by two-temperature heating dynamics.
The results of two-temperature model (TTM) 2D simulations are presented on the irradiation of gold by a single 800~nm femtosecond laser pulse whose intensity is modulated in order to reproduce a small initial temperature perturbation, which can arise from incoming and scattered surface wave interference. The growing (unstable) modes of the  temperature distribution along the surface may be responsible for the LIPSS (Laser Induced Periodic Surface Structures) formation.
After the end of the laser pulse and before the complete coupling between lattice and electrons occurs, the evolution of the amplitude of the subsequent modulation in the lattice temperature reveals different tendencies depending on the spatial period of the initial modulation. 

This instability-like behaviour is shown to arise due to the perturbation of the electronic temperature which relaxes slower for bigger spatial periods and thus imparts more significant modulations to the lattice temperature. Small spatial periods of the order of 100 nm and smaller experience stabilization and fast decay from the more efficient lateral heat diffusion which facilitates the relaxation of the electronic temperature amplitude due to in-depth diffusion.
An analytical instability analysis of a simplified version of the TTM set of equations supports the lattice temperature modulation behaviour obtained in the simulations and reveals that in-depth diffusion length is a determining parameter in the dispersion relation of unstable modes.
Finally it is discussed how the change in optical properties can intensify the modulation-related effects.

\end{abstract}



\maketitle

\section{1. Introduction}

Short and ultrashort laser pulses are known to generate LIPSS (Laser Induced Periodic Surface Structures) or ripples on the surfaces of metals, dielectrics and semiconductors. The pattern appears if the surface is exposed to multiple \cite{SipeGe,Vorobyev} or even single \cite{MyPRE, SPRPF, Raciukaitis_2015} laser shots. Although the LIPSS have been observed from 1960s on laser-irradiated surfaces of materials of different kind, mechanisms of their formation are still controversially discussed in the literature. The first observation and explanation of the periodic pattern induced by ruby laser ablation of semiconductors was reported by Birnbaum \cite{Birnbaum}; he explained the pattern by the diffraction of the laser beam on optical elements of the set up. Nowadays the most common approach to the explanation of the patterns is based on interference of the incident laser beam with the 
surface-scattered wave
\cite{Akhmanov,Sipe,Bonse2009,Garrelie2011}. The resulting periodic pattern of the energy deposited at the target surface manifests itself in the periodically-patterned material ablation. In the frames of the plasmonic model, the excited surface wave is a surface plasmon polariton (SPP), and the standing wave is formed by the interference of the plasmonic and laser fields, which results in a spatially modulated electron heating along the surface. 

The surface scattered wave and plasmonic models can explain the period and the orientation of the patterns under certain irradiation conditions, accounting for ablation from the peaks of the periodic absorption pattern. However, many contradictions between these models and experimental observations also exist in the literature \cite{AcuosticWave, Clark,SPRPF}. These contradictions raise the question on whether other mechanisms/processes can contribute to the formation of the diversity of the observed periodic surface structures, especially during post-irradiation evolution of laser-disturbed surfaces. One of the most obvious candidates for the alternative mechanism of the LIPSS formation would be a hydrodynamic instability of the liquid melt on the surface. Indeed, there are many processes, which can cause periodic pattern formation \cite{Oron,LIPSSroughness,gurevich_hydro_2016_ASS} in thin liquid layers.
In \cite{Varlamova}, a phenomenological  approach based on the Kuramoto-Sivashinsky model \cite{Cross} was proposed. It is capable to describe the ripple formation in different systems such as ion beam sputtering, water jet cutting \cite{Friedrich1,Friedrich2}. A similarity between these processes and the femtosecond laser ablation validates this approach. In the phenomenological Kuramoto-Sivashinsky-based model developed in \cite{Varlamova}, the authors suggest a positive feedback mechanism based on the different desorption velocities of atoms on a rough surface, provided that the surface layer of the metal is charged. 
The model describes the morphology of the structures for multiple pulse ablation but cannot give an explanation for single-pulse pattern formation. Furthermore, it does not predict the experimental parameters (laser pulse characteristics, material properties), which lead to the periodic pattern formation and the period of the self-organized structures.

Although phenomenological models enable a convincing description of bifurcation scenarios, they cannot answer the question how the physical parameters of the system, which can be controlled in experiments, are related to the coefficients introduced into such models as bifurcation parameters. As a result, predictions made in the frames of these theories can not be directly verified experimentally. However, the most of the physical processes upon femtosecond laser ablation of metals are well-understood and some of them can be described analytically, suggesting the possibility of developing a realistic physical model of the LIPSS formation.

\begin{figure}
 \includegraphics[width=8cm]{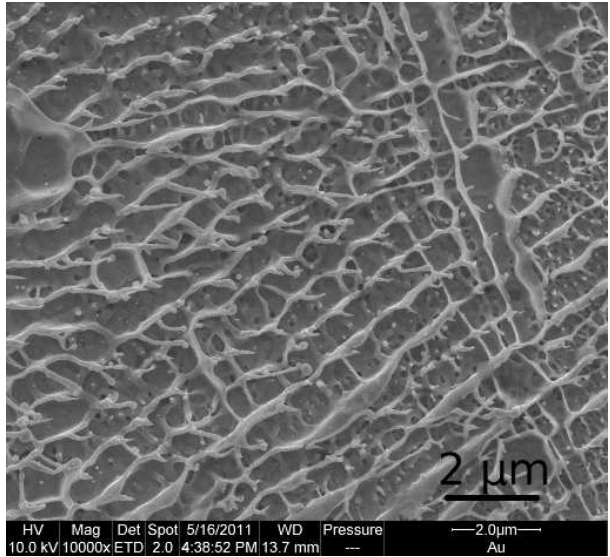}
\caption{Electron microscope image of a gold surface exposed to single fs laser pulse. Normal incidence, peak fluence $F_0\approx 4\,\mathrm{J\cdot cm^{-2}}$, $\lambda_{\textrm{las}} = 800$\,nm and $\tau_p \approx 100$\,fs.}
 \label{LipssAu}
\end{figure}

The physical processes triggered on the surface of metals by ultrashort laser pulses can be treated in frames of the two-temperature model (TTM), whose details are described in many publications, see e.g~\cite{Anisimov,corkum,levy_relaxation_2016_ASS}. In this work, a hydrodynamic-like approach is proposed to explain LIPSS observed on metallic surfaces upon single femtosecond laser pulse irradiation, see Fig.~\ref{LipssAu}, and the analytical results are compared with numerical calculations based on the full 2D TTM with temperature-dependent material properties \cite{Lin}. We simulate numerically the interaction of femtosecond laser pulses at the wavelength $\lambda_{\textrm{las}}=800$\,nm with a gold sample and introduce a single mode perturbation of different spatial period into the laser intensity profile. Then the evolution of the amplitude of the subsequent modulation in the lattice temperature is analyzed both numerically and analytically.
The paper is organized as follows. In Section 2, the numerical model and its results are presented. The numerical results are supported by the analytical stability analysis of a simplified TTM set of equations whose details are given in Section 3. In Section 4, the application of this study to the regimes of the LIPSS formation are discussed with addressing a potential contribution of the change in optical properties due to swift laser heating of the electron subsystem. 

\section{2. Numerical Analysis}\label{SEC_NumAnalysis}

\subsection{Modelling}

The description of the interaction between a femtosecond laser pulse and a gold sample is based on the TTM which provides the dynamics of the electron and lattice temperatures, $T_e$ and $T_l$ respectively~\cite{Anisimov}. The simulations have been performed in 2D, in the $(x,z)$-plane for a gold sample. The coordinate $z$ points in the direction of the laser beam propagation and $x$ in the transverse direction, parallel to the sample surface; incidence is normal, see Fig.~\ref{FIG_IUnifo_SurfTeTl_t}~(a). The model in its full form is written as
\begin{equation}\label{TTMeq}
 \left\{
  \begin{aligned}
     c_e\partial_t T_e=&\partial_z\kappa_e\partial_z T_e + \partial_x\kappa_e\partial_xT_e -G\big(T_e-T_l\big) +Q(x,z,t)\\
     c_l\partial_tT_l=& \kappa_l\left(\partial^2_z+\partial^2_x\right) T_l +G\big(T_e-T_l\big)
  \end{aligned}
\right.
\end{equation}

Here the indexes $e$ and $l$ are for electrons and lattice respectively, $c$ is the specific heat capacity, $\kappa$ - the specific thermal conductivity and $G$ - the electron-phonon coupling factor. The coupling factor, electronic heat capacity and thermal conductivity are considered as in Ref.~\cite{wellershoff_role_1999_APA}: $c_e = A_e T_e$ and $\kappa_e =  \varkappa \frac{T_e}{T_l}$ with fixed $A_e$ and $\varkappa$ (all parameters are summarised in Table~\ref{TAB_GoldOpticThermalCoef}). The lattice thermophysical properties are considered to be constant in this study.
The last term in the equation for the electron temperature
\begin{equation}\label{EQ_TTMSourceTerm}
Q(x,z,t)=\mathcal{W}(x,t)(1-R)\alpha \exp({-\alpha z}) 
\end{equation}
accounts for the absorption of the energy of the laser pulse whose spatio-temporal shape and amplitude are described by the function $\mathcal{W}(x,t)$. 
The incidence of the laser is considered to be normal to the gold sample and the latter to present an ideally smooth surface, characterized by the reflection and absorption coefficients $R$ and $\alpha$ respectively. The absorption coefficient $\alpha = 4\pi n'' / \lambda_\textrm{las}$, is the inverse of the optical depth in the metal where $n''$ is the extinction coefficient of the complex index of refraction $n = n' + i n''$. The temporal shape of the pulse is considered to be Gaussian with duration $\tau_p=100$~fs (FWHM). In the transverse direction, the incident laser intensity is considered to be either uniform, $f(x)\equiv 1$, or with variations, e.g. intensity modulation resulting from the interference of the incident and surface-scattered light:
\begin{equation}\label{EQ_TTMSourceShape}
\mathcal{W}(x,t) = f(x) \times \fracd{2F_{0}}{\tau_{p}}\sqrt{\fracd{\ln2}{\pi}}\exp \left( -4\ln2\ \fracd{(t-2\tau_p)^{2}}{\tau_{p}^{2}} \right)
\end{equation}
The time moment $t=t_\textrm{ini}=0$ corresponds to the beginning of the simulations, when the laser pulse starts. The laser pulse intensity is maximum at $t=2\tau_{p}$ and the pulse terminates at $t_0=4\tau_{p}$.
In this Section, a simplified case of the absorbed laser fluence $F_0 = F_\textrm{abs}$, is considered, disregarding the influence of the dynamic change of reflectivity (considered in Section 4).
The set of equations~\eqref{TTMeq}, \eqref{EQ_TTMSourceTerm} and \eqref{EQ_TTMSourceShape} is solved numerically using a 2D finite difference implicit scheme with splitting by the directions and applying the Thomas algorithm~\cite{godunov_differences_1987}. Initially, the temperatures of electrons and lattice are equal and uniform over the sample, $T_{e}(t_\textrm{ini}) = T_{l}(t_\textrm{ini}) = 300\units{K}$. No-flux conditions are set at all boundaries. The thickness of the numerical region in the gold sample is chosen sufficiently deep to avoid heat flux reflection from the remote surface.

\begin{table}[h]								
\begin{centering}								
\begin{tabular}{|c|c|c|c|c|}							
\hline								
 {\small Parameter} 	&	 {\small Symbol} 	&	 {\small Value}  	&	 {\small Unit} 	&	 {\small Reference} \\
\hline								
{\small Lin. coef. of elec. heat capacity}  	&	 {\small $A_e $}  	&	  {\small 71 } 	&	 {\small $\mathrm{J/(m^3\cdot K^2)}$} 	&	 {\small $c_e = A_e \ T_e$,  \cite{wellershoff_role_1999_APA}}  \\
{\small Lin. coef. of elec. therm. cond.}  	&	 {\small $\varkappa$}  	&	  {\small 318 } 	&	 {\small $\mathrm{W/(m\cdot K)}$} 	&	 {\small $\kappa_e = \varkappa \ T_e/T_l$,  \cite{wellershoff_role_1999_APA}}  \\
{\small Coupling factor}  	&	 {\small $G$}  	&	  {\small $2.1\E{16}$ } 	&	 {\small $\mathrm{W/(m^3\cdot K)}$} 	&	 {\small \cite{wellershoff_role_1999_APA}}  \\
{\small Lattice heat capacity}  	&	 {\small $c_l$}  	&	  {\small $2.5\E{6}$} 	&	 {\small $\mathrm{J/(m^3\cdot K)}$} 	&	 {\small \cite{wellershoff_role_1999_APA}}  \\
{\small Lattice therm. conductivity}  	&	 {\small $\kappa_l$}  	&	  {\small  $1$ } 	&	 {\small $\mathrm{W/(m\cdot K)}$} 	&	 \\
 {\small Refractive index, 800 nm}  	&	 {\small $n'$}  	&	 {\small 0.181} 	&	 {\small $-$} 	&	 {\small \cite{Palik1985}} \\
 {\small Extinction coefficient, 800 nm}  	&	 {\small $n''$}  	&	 {\small 5.12}  	&	 {\small $-$}  	&	 {\small \cite{Palik1985}} \\
\hline								
\end{tabular}								
\par\end{centering}								
\caption{Thermophysical and optical parameters used in the numerical simulations of laser-irradiated gold.}~\label{TAB_GoldOpticThermalCoef} 					
\end{table}

\subsection{Simulations results}

\begin{figure}
\begin{center}
\includegraphics[width=0.35\textwidth]{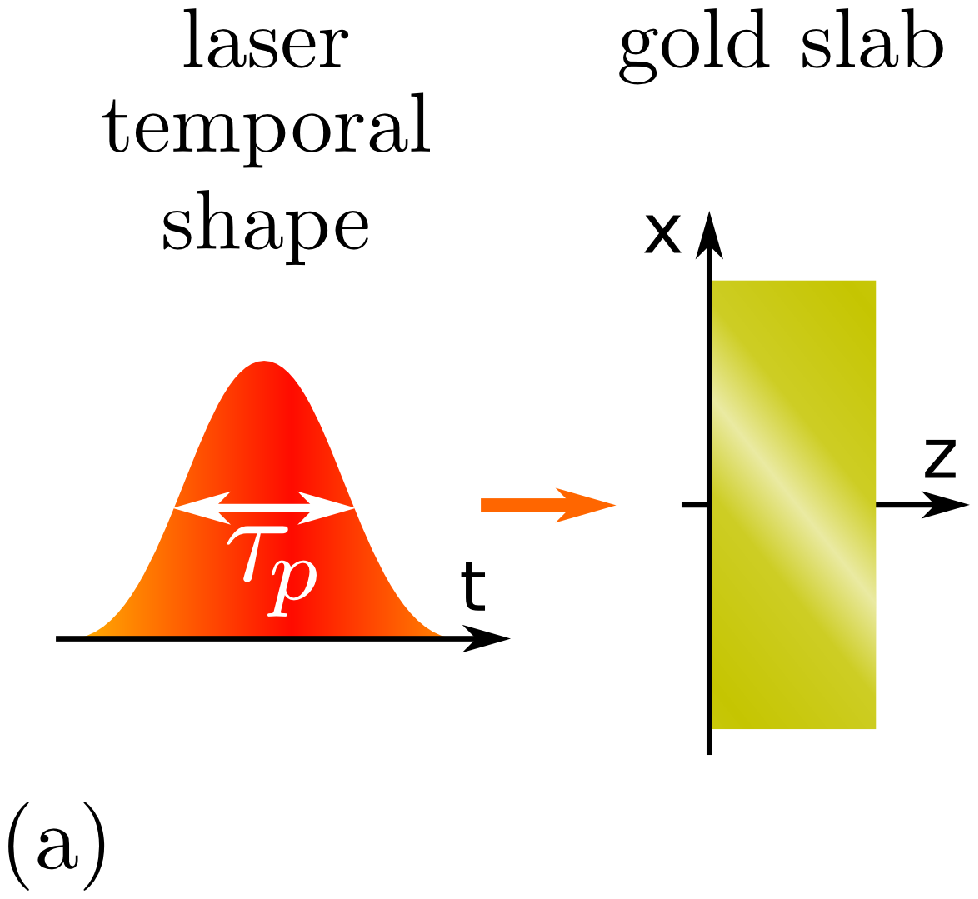}\hspace*{1cm}
\includegraphics[scale=0.97]{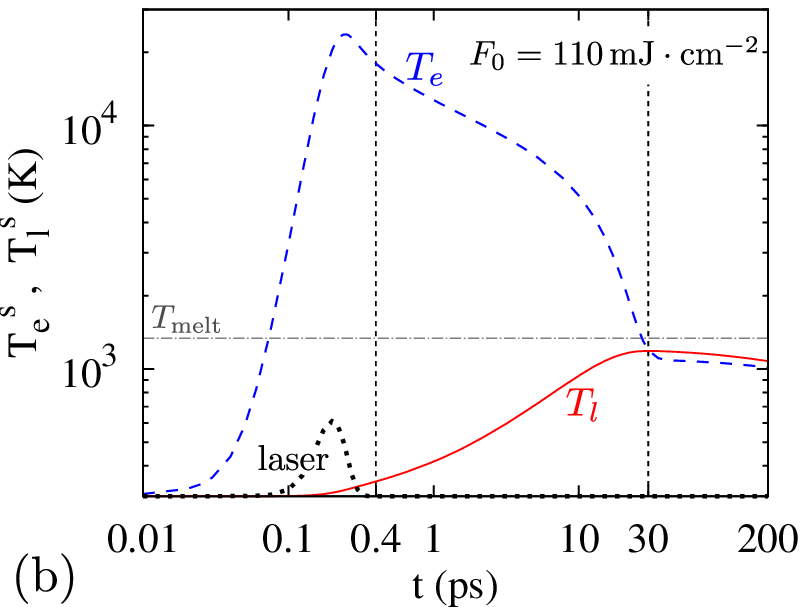}
\end{center}
\caption{(a) Sketch of the numerical modeling geometry. (b) Evolution of the electron (dashed line) and lattice (solid line) temperatures on the surface of the gold slab. The laser pulse is shown by dotted line (arbitrary units). The melting temperature is outlined by horizontal dot-dashed line. The time moments of the laser pulse termination $t_0=0.4\,\mathrm{ps}$ and electron-lattice thermalization $t_c\approx 30\,\mathrm{ps}$ are marked by vertical dashed lines. Absorbed peak fluence is $F_0 = 110\,\mathrm{J\cdot cm^{-2}}$.}\label{FIG_IUnifo_SurfTeTl_t}
\end{figure}

First we present the simulation results for one-dimensional case ($f(x)=1$ in Eq.~\eqref{EQ_TTMSourceShape}). The evolution of the electron and lattice temperatures on the gold surface upon laser irradiation is shown in Fig.~\ref{FIG_IUnifo_SurfTeTl_t} for the absorbed peak fluence of $F_{0} = 110\units{mJ\cdot cm^{-2}}$. The electron temperature is swiftly increasing during the first part of the laser pulse action. After the laser pulse termination, it starts to decrease due to both heat diffusion towards the material bulk and the electron-lattice relaxation through electron-phonon collisions characterized by the coupling factor $G$. The lattice temperature, whose increase is conditioned only by these collisions, slowly rises until the two subsystems reach similar temperatures after a coupling time of $t_c \sim 30\,\mathrm{ps}$.

%
%


\begin{table}
\begin{center}
\begin{tabular}{|c|c|c|}
\hline
Numbering	&	$\Lambda$ ($\mathrm{\mu m}$) 	&  $k$ ($\mathrm{\mu m^{-1}}$)	\\
\hline
1			&	0.08	 	&	78.5		\\
2			&	0.30	 	&	20.9		\\
3			&	0.80	 	&	7.85		\\
4			&	1.20		&	5.24		\\
5			&	2.10		&	2.99		\\
6			&	3.00		&	2.09		\\
7			&	8.00		&	0.785	\\
\hline
\end{tabular}
\end{center}
\caption{Perturbation modes modeled in this paper.}\label{TAB_ModeNumbering}
\end{table}

Then, a series of calculations have been performed on the irradiation of the gold samples with a modulated laser intensity profile along the surface, similar to \cite{levy_relaxation_2016_ASS}. The modulation is implemented by introducing into~\eqref{EQ_TTMSourceShape} the following term:
\begin{equation}\label{EQ_TransverseModulations}
f(x) = \left( 1 + \eta \cos\left( kx\right) \right).
\end{equation}
Different single modes of harmonic perturbation, $k=2\pi/\Lambda$, are thus compared, where $\Lambda$ is the perturbation spatial period. The values of $k$ used in simulations are summarized in Table~\ref{TAB_ModeNumbering}. For all simulations the initial perturbation amplitude was set to $\eta = 0.05$.
From the physical point of view, this modulation models an initial perturbation of the laser energy deposition, which can be of different origins. One of the possible reasons of this modulation can be interference of the incoming laser pulse with the scattered electromagnetic wave induced by surface roughness or with excited SPP~\cite{Sipe}. Another reason can be inhomogeneities in the surface absorption due to, e.g., surface defects, scratches, impurities or nanoparticles present on real surfaces. Here we have chosen the harmonic profile, which enables to study effects of the spectral components of the temperature perturbation. Its invariance along the non-modelled $y$ coordinate assumes a geometry favourable to the development of surface ripples with ridges oriented along this direction.

\begin{figure}
\begin{center}
\includegraphics[scale=0.98]{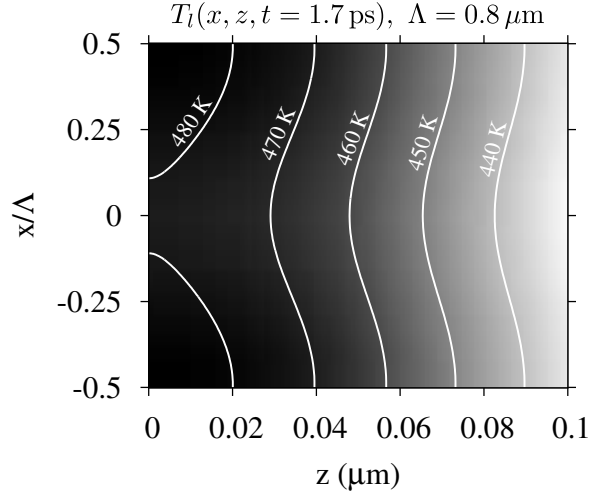}
\end{center}
\caption{2D distribution of the gold lattice temperature with isolines at~$t=1.7$\,ps. The laser pulse comes from the left. Darker areas correspond to higher lattice temperature. Only one perturbation period along the surface and the first 100\,nm from the surface toward the bulk are shown. Laser parameters are the same as for Fig.~\ref{FIG_IUnifo_SurfTeTl_t}. The intensity modulation period is $\Lambda = 0.8\,\mathrm{\mu m}$.}\label{Tl_2Dplot}
\end{figure}

During the irradiation, the perturbation in the laser intensity profile results in the spatial modulation of the electron temperature and finally affects the lattice. An example of the modulated lattice temperature distribution in the sample at $t=1.7$\,ps is shown in Fig.~\ref{Tl_2Dplot} for $F_0=110\,\mathrm{J\cdot cm^{-2}}$. The figure presents the region of one modulation period along the surface and 100\,nm thick toward the bulk. Isotemperature contours highlight the modulation amplitude, which is defined along the surface as $a_l(t)=T_l^{max}(t)-T_l^{min}(t)$ with $T_l^{max}$ and $T_l^{min}$ to be the maximum and minimum lattice temperature on the surface respectively. The modulation of the electron temperature along the surface, $a_e(t)$, is similarly defined. 

\begin{figure}
\begin{center}
\includegraphics[scale=0.97]{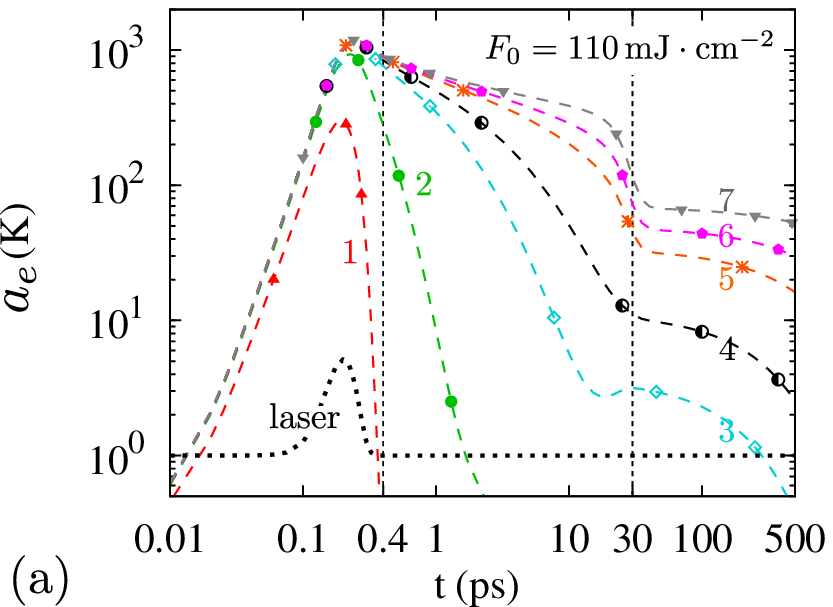}
\hfill \includegraphics[scale=0.97]{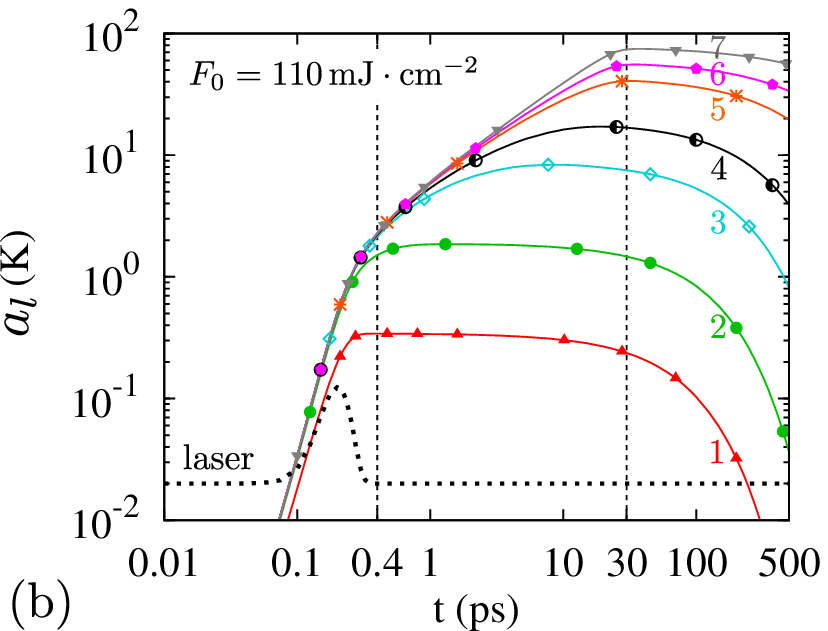}
\end{center}
\caption{Amplitude of the modulations of the electron~(a) and lattice~(b) temperatures on the surface of gold for different modulation modes (numbers correspond to those in Table~\ref{TAB_ModeNumbering}). Simulations were performed with parameters summarized in Table~\ref{TAB_GoldOpticThermalCoef}}\label{FIG_AmpliTlModul}
\end{figure}

Figure~\ref{FIG_AmpliTlModul} shows the temporal evolution of the amplitudes $a_e(t)$ and $a_l(t)$. During the laser pulse action, the effects of electron heat diffusivity are weak. Therefore, the absorbed laser energy accumulated during the pulse is higher in the areas irradiated by the peaks of the spatially modulated intensity profile compared to that of the intensity minima. This results in the growth of the electron temperature modulation along the surface, $a_e(t)$. 
However, as can be seen from Fig.~\ref{FIG_AmpliTlModul}, the maximum of $a_e(t)$ is reached at slightly different time moments for the different modes and the modulation amplitude starts to decrease already before $t_0$. This indicates that the electron heat diffusion comes into play already before the laser pulse termination and it has a higher importance for smaller modulation periods (i.e., higher temperature gradients). After the end of the laser pulse, the amplitude of the electron temperature modulation is gradually decreasing due to heat dissipation and, to a smaller extent, due to electron-lattice coupling.

\begin{figure}
\begin{center}
\includegraphics[scale=0.98]{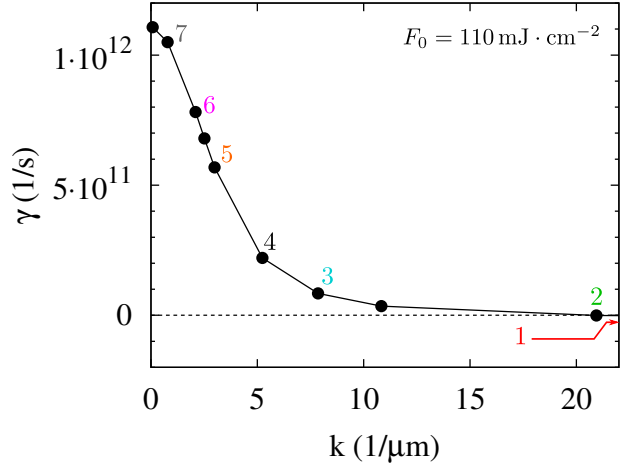}
\end{center}
\caption{Dispersion relation obtained from the averaged growth rate of single mode simulations performed with the same parameters as in Fig~\ref{FIG_IUnifo_SurfTeTl_t}. Numbering corresponds to the modes listed in Table~\ref{TAB_ModeNumbering}. The cut-off for the growth of the modulation amplitude after the laser pulse termination is $k = k_\textrm{c} \sim 20\,\mathrm{\mu m^{-1}}$.
}\label{GrowthRates}
\end{figure}

Since, at the considered laser fluence, the electrons are heated to a high temperature exceeding 2 eV (Fig.~\ref{FIG_IUnifo_SurfTeTl_t}), the amplitude of the lattice temperature modulation, $a_l(t)$, starts to increase already during the laser pulse due the coupling between the electrons and the lattice. Interestingly, after the laser pulse termination, $t_0$, and up to the coupling time $t_c$, the amplitudes of the lattice temperature modulation of several modes continue to grow whereas the amplitudes of others gradually decrease. This transient growth concerns the smallest wave numbers $ k\lesssim 20\,\mathrm{\mu m^{-1}}$ (i.e. spatial periods $\Lambda\gtrsim 0.3\,\mathrm{\mu m}$). The growth rates averaged over the period $[t_0;t_c]$ 
and defined as $\gamma=\left(a_l(t_c)-a_l(t_0)\right)/\left( (t_c-t_0) \, a_l(t_0) \right)$ for different wave numbers, are summarized in Fig.~\ref{GrowthRates}. It is seen that the cut-off for growth of the modulation amplitude after the laser pulse termination is located at $k_\textrm{c} \sim 20\,\mathrm{\mu m^{-1}}$ and modes with spatial periods smaller than~$\sim300$\,nm are stable.

We underline that all modes yield the same average surface temperature evolution (for both electrons and lattice) and similar thermalization times $t_c=30\pm2$\,ps for the considered laser fluence
(the deviation, for a single mode perturbation, comes from the difference between the regions of temperature minima and the regions of temperature maxima, due to the temperature-dependent electron-lattice coupling rate). 
Only the evolution of the modulation amplitudes $a_e(t)$ and $a_l(t)$ differs significantly for different modes. Furthermore, the same mode-dependent behaviour is observed for the evolution of the amplitudes of the lattice and electron temperatures when applying a smaller ($\eta' = 0.01$) or a higher ($\eta'' = 0.1$) initial amplitude of intensity modulation in Eq.~\eqref{EQ_TransverseModulations}. Respectively, the reached amplitudes of $T_e$ and $T_l$ modulations are smaller or higher but the evolution of the amplitudes normalized to their values at the end of the laser pulse is identical. In this sense, the evolution of the perturbations of the temperature along the surface presents a transient instability-like behaviour with a spectrum of unstable modes corresponding to spatial periods above~$\sim 0.3\units{\mu m}$. Notice that the period of the experimentally observed pattern, $\Lambda_{\textrm{exp}} \sim 800\,\mathrm{nm}$, see Fig.~\ref{LipssAu}, is within this range.

\section{3. Analytical analysis}

On the qualitative level the following mechanism of the instability-like behavior can be suggested.
First, since the lattice heat diffusion is much smaller than that of the electrons, up to the electron-lattice thermalization time, the modulations received from the electron subsystem are accumulated and almost ``freezing" in the lattice. In the simulations presented, the electron temperature modulation is then the main source of the evolution of the lattice temperature modulation until their amplitude reach a similar value. Therefore the amplitude of the lattice temperature grows with a rate that depends on the difference $a_e(t)-a_l(t)$. 
Secondly, the main cause that heals the electron temperature perturbation is the heat diffusion in the transverse direction (along the surface). For smaller spatial periods, the transverse gradient of temperature is higher and, hence, the time for smoothing modulations is shorter, causing a fast decrease as shown in the Fig.~\ref{FIG_AmpliTlModul}~(a). As a result, the small-period lattice modulations have a limited time to develop and exhibit low final amplitudes.
The modulation smoothing time, which can be estimated as $\tau_{e} \sim \left(\frac{\Lambda}{2}\right)^2 {D}_e^{-1}$, scales with the squared perturbation period. It varies from hundreds of fs for the highest studied mode, $k=78.5\,\mathrm{\mu m^{-1}}$, to more than one~ns for $k=0.785\,\mathrm{\mu m^{-1}}$. 
Finally, when the lattice temperature modulation becomes significant, the smoothing effect is diminished by different efficiency of the heat transport towards the bulk. For a given mode, the regions of the crests of the electron temperature exhibit lower electron heat diffusivity $D_e \sim \kappa_e/ C_e \sim \varkappa/(A_e T_l)$ than in the troughs because, due to the electron-lattice coupling, they also correspond to the regions of crests of the lattice temperature. Therefore the electron subsystem is more significantly cooled in the trough regions than in the crests, by the heat transfer towards the bulk.

In order to deepen the understanding of the conditions at which the instability develops and on how these conditions depend on the material properties, an analytical model has been developed.
It is based on the system of equations~\eqref{TTMeq} without the laser source term, $Q(x,z,t)$, as we are interested only in the post-irradiation evolution of the electron and lattice temperatures of a gold sample exposed to a spatially-modulated laser irradiation as considered in Section 2.

The structure of the system~\eqref{TTMeq} is similar to that of the reaction-diffusion equations, which describe pattern formation phenomena, e.g. in chemical reactions, electrical discharges, and on the skin of some animals~\cite{Turing, Cross,Murray,Charge}. Here the procedure similar to that provided in~\cite{strogatz} for solving the reaction-diffusion system is applied to the TTM in order to analyse whether it can be unstable with respect to small periodic fluctuations of temperature in the plane of the sample surface.

The behaviour of the system~(\ref{TTMeq}) is usually analysed in the $(t,z)$ plane, i.e., it is used to study the evolution of the in-depth profile of the temperatures. In this work, in order to explain the pattern formation in the plane of the sample surface, i.e., along the $x$ coordinate, the problem is divided into 2 steps. First, it is assumed that the temperature profile variation in the surface plane is negligible compared to that along the $z$ axis. Hence, the one-dimensional problem of laser-induced heating of a gold sample can be solved numerically with temperature dependent material parameters. This allows to estimate the in-depth profile characteristic length $\xi^{-1}(t)$ to be defined below. After that, the perturbation method is applied to evaluate analytically the stability of the system~(\ref{TTMeq}) in respect of the small perturbations of the electron and lattice temperatures on the surface and derive the instability criterion with the obtained in-depth length.

\begin{figure}
 \includegraphics[scale=0.98]{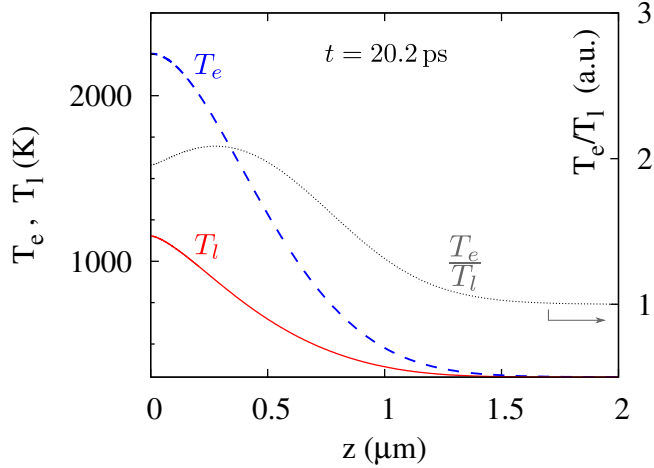}
\caption{Calculated in-depth profile of the electron (dashed line) and the lattice (solid line) temperatures in a gold sample at $t=20.2$\,ps. The laser parameters used are the same as in Fig.~\ref{FIG_IUnifo_SurfTeTl_t}. 
Thin dotted line is the ratio between the electron and lattice temperatures (right axis).}
 \label{FigTTM}
\end{figure}

\paragraph{Step 1:} Homogeneous temperature distributions $T_{j}(x,z,t)=T_{j}(z,t)$, the index $j$ denotes the electrons and lattice components ($j$ = $e$,$l$) are assumed along $x$-direction and the system~\eqref{TTMeq} is solved numerically with the parameters listed in Table~\ref{TAB_GoldOpticThermalCoef}. In Fig.~\ref{FigTTM}, the temperature in-depth profiles are shown for the electrons and lattice in gold for the time moment $t=20.2$\,\textrm{ps}, calculated with the TTM numerical tool developed for metals and presented in Ref.~\cite{levy_relaxation_2016_ASS}. The irradiation parameters are the same as in Figs.~\ref{FIG_IUnifo_SurfTeTl_t},~\ref{Tl_2Dplot} and~\ref{FIG_AmpliTlModul}.

The profile of the electron temperature can be approximated by the function $T_{e}(z)\propto{}\exp(-\xi'z^2)$~\cite{corkum}, somewhat distorted by the electron-phonon coupling and by the temperature dependent coefficients of the system \eqref{TTMeq}. The analytical description of the in-depth profile of the lattice temperature can be even more complicated in the case of melting due to the plateau of constant temperature between the molten and the solid material where the heat of fusion is consumed by the lattice. For simplicity in the following analytical treatment, the profiles are considered in the form of an exponential decay with the characteristic length scale $\xi^{-1}$: $T_{j}(z,t)\propto T_{j}^0(t)\exp(-\xi z)$. 
From the numerically obtained in-depth profiles of the electron temperature for the case of uniform irradiation, this characteristic length is estimated as the depth at which the temperature drops from its surface value, $T_{e}^0$, to $\left( T_{e}^0 - 300 \right)/\exp(1)$.
Between the end of the laser pulse and the electron-lattice thermalization time, at $F_0=110\,\mathrm{mJ\cdot cm^{-2}}$, the characteristic length (globally increasing in time) ranges from $\xi^{-1}\sim 0.15\,\mathrm{\mu m}$ to $\xi^{-1}\sim 0.7\,\mathrm{\mu m}$.



\paragraph{Step 2:} We assume the periodic modulation of the surface temperature as a small perturbation of the homogeneous solution $T_{j}^s(x,t)=T_{j}^0(t)+\tilde{T}_{j}(t)\!\cdot\!e^{ikx}$, $j$ = $e$,$l$. Here $\tilde{T}_{j}(t)$ is the small temperature perturbation amplitude and $k=2\pi/\Lambda$ is the wave vector's amplitude of the perturbation ($\Lambda$ is the period). This amplitude of the temperature modulation is related to that introduced in Section~2 as $\tilde{T}_{j}=a_{j}/2$. Then $T_{j}(x,z,t)=T_{j}^s(x,t)\exp(-\xi z)$ is substituted into the system \eqref{TTMeq} and the system behaviour is analysed after the laser pulse termination. The last assumption allows us to neglect the source term in the equation for the electron temperature. The simplified, linearized system for the evolution of the surface temperature perturbations \eqref{TTMeq} can be written as
   \begin{equation}
 \left\{
   \begin{aligned}
      A_eT^0_ee^{-\xi z}\myfrac{\mathrm{d} \tilde{T}_e}{\mathrm{d} t}=& \varkappa\xi^2\myfrac{{T^0_e}^2}{T^0_l}\left(2\myfrac{\tilde{T}_e}{T^0_e}-\myfrac{\tilde{T}_l}{T^0_l}\right) -k^2\varkappa\myfrac{T^0_e}{T^0_l}\tilde{T}_e-G(\tilde{T}_e-\tilde{T}_l)\\
      c_l\myfrac{\mathrm{d} \tilde{T}_l}{\mathrm{d} t }=& \kappa_l\xi^2 \tilde{T}_l - \kappa_lk^2\tilde{T}_l+G(\tilde{T}_e-\tilde{T}_l)
   \end{aligned}
 \right.
 \label{sysTsurf}  
   \end{equation}

On the sample surface ($z=0$) this system can be written in the matrix form $\dot{\mathbf{w}}=\mathbb{A}\mathbf{w}$, where dot denotes the time derivative and the components of the vector $\mathbf{w}$ are the amplitudes of the temperature perturbations of the electrons and the lattice.
The coefficients of the matrix $\mathbb{A}(k)$  are listed in Table~\ref{TabParam}. They depend on the mode number $k$ of the perturbation and can be calculated with known material properties of the processed metal. 

\begin{table}
 \caption{Coefficients of the matrix $\mathbb{A}(k)$ corresponding to the system of equations \eqref{sysTsurf}}\label{TabParam}
\begin{tabular}{|c|c|}
 \hline
 Coefficient & Equation \\
 \hline
$a_{11}$ & $\myfrac{2\varkappa\xi^2}{A_eT^0_l}-\myfrac{k^2\varkappa}{A_eT^0_l}-\myfrac{G}{A_eT^0_e} $\\
$a_{12}$ & $-\myfrac{\varkappa T^0_e\xi^2}{A_e{T^0_l}^2}+\myfrac{G}{A_eT^0_e}$\\
$a_{21}$ & $\myfrac{G}{c_l}$ \\
$a_{22}$ & $\myfrac{\kappa_l\xi^2}{c_l}-\myfrac{k^2\kappa_l}{c_l}-\myfrac{G}{c_l}$\\
\hline
\end{tabular}
\end{table}

\subsection{Stability analysis}

A small perturbation of the surface temperature of a mode $k$ will grow if the matrix $\mathbb{A}(k)$ defined in Table~\ref{TabParam} is unstable \cite{strogatz}. Instead of calculating the eigenvalues of the matrix numerically, we can first find the analytical equation for the parameters, at which the matrix looses the stability. These parameters can be found by means of the trace-determinant criterion \cite{strogatz}: the matrix is unstable when either the trace of the matrix is positive $\mathfrak{Tr}(\mathbb{A}(k))=a_{11}+a_{22}>0$ or the determinant is negative $\mathfrak{Det}(\mathbb{A}(k))=a_{11}a_{22}-a_{12}a_{21}<0$. When considering the coefficients from Table~\ref{TabParam} and noting that $\kappa_l/c_l\ll \varkappa/A_eT^0_e$, the following instability criterion for the mode $k$ can be derived from the condition of the positive trace of the matrix:
\begin{equation}\label{EQ_UnstableAnalyticalCond}
k^2<2\xi^2-\frac{G}{\varkappa}\frac{T^0_l}{T^0_e}=k^2_c.
\end{equation}
The analysis of the second part of the criterion $\mathfrak{Det}(\mathbb{A}(k))<0$ is more elaborate. However, it can be simplified by restricting consideration to the case $k>k_c$. Indeed, if $k<k_c$, the matrix is already unstable according the trace-criterion in the equation~\eqref{EQ_UnstableAnalyticalCond} so that the determinant-criterion does not bring any new information. It can be shown that for $k>k_c$ the coefficients $a_{11}$, $a_{22}$ and $a_{12}$ are negative and, consequently, the determinant is always positive. Hence, this criterion does not influence the analytical stability analysis.

Equation~\eqref{EQ_UnstableAnalyticalCond} defines the band of unstable wave numbers of the periodic temperature modulation. During first picoseconds after the laser pulse has excited the surface, $T_e\gg T_l$ and a broad band of unstable modes, $k\lesssim \sqrt{2}\xi$, starts to grow. Perturbations with small $k$ (i.e, large periods) are always unstable, whereas the modes of smaller periods are stabilized by the lateral diffusion as explained in the beginning of the current section. For example, for $T_e\approx 1.8~10^4\,\mathrm{K}$ and $T_l\approx 350\,\mathrm{K}$, which correspond to the end of the laser pulse (see Fig.~\ref{FIG_IUnifo_SurfTeTl_t}), the estimation of the critical wavelength according to Eq.~\eqref{EQ_UnstableAnalyticalCond} gives $k_c\approx 9.4\,\mathrm{\mu m}^{-1}$ for $\xi^{-1}=0.15\,\mathrm{\mu m}$. This is lower than the cut-off obtained by modeling (Fig.~\ref{GrowthRates}) but provides the correct order of magnitude.

When the difference between $T_e$ and $T_l$ decreases due to gradual electron-lattice coupling process, the modes, which were unstable while the system was far from thermal equilibrium, become stable and the amplitude of the temperature modulation starts to decrease. This scenario follows from Eq.~\eqref{EQ_UnstableAnalyticalCond} and is confirmed by numerical simulations (Fig.~\ref{FIG_AmpliTlModul}). The amplitudes of the temperature modulation corresponding to larger periods (smaller $k$) have more time to grow.

\begin{figure}
 \includegraphics[width=8cm]{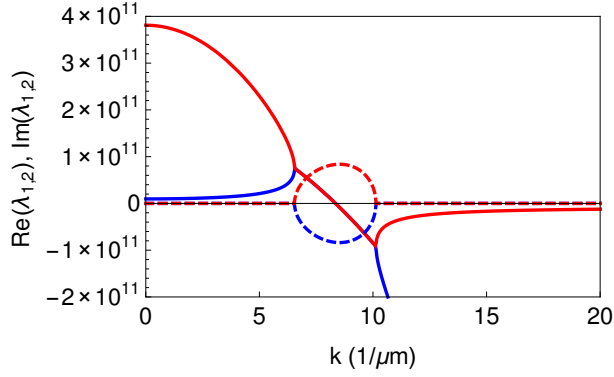}
 \caption{Two eigenvalues of the system~\eqref{TTMeq} calculated for Au with $\xi^{-1}=0.15\,\mathrm{\mu m}$. Solid line - the real part, dashed line - imaginary part of the eigenvalues; blue and red lines correspond to different eigenvalues.}
\label{FigEW}
\end{figure}

Here we stress that the temperature modulation does not disappear immediately as soon as the system \eqref{TTMeq} becomes stable. The amplitude starts to decrease but the modulation needs some time to vanish. The time scale of the instability is defined by the maximal eigenvalue $\lambda_1$ or $\lambda_2$ of the matrix $\mathbb{A}(k)$, because the evolution of the amplitude of a small instability in the linear case is described as $\mathbf{A}(t)=\mathbf{A}_{01}\exp({\lambda_1t})+\mathbf{A}_{02}\exp({\lambda_2t})$ with $\mathbf{A}_{01}$ and $\mathbf{A}_{02}$ to be initial values of the perturbation amplitude. The eigenvalues of the matrix $\mathbb{A}(k)$ are calculated numerically and plotted in Fig.~\ref{FigEW} for different wave numbers $k$ and fixed $\xi^{-1}=0.15\,\mathrm{\mu m}$

Both eigenvalues are negative for large $k$, and positive for $k<k_c$. It is interesting that, close to $k_c$, the eigenvalues have nonzero imaginary parts, which should induce oscillations of the temperature perturbation in time. This is known as the Hopf bifurcation. The band of wave numbers, for which the Hopf bifurcation occurs, can be calculated analytically from the condition $\mathfrak{Tr}(\mathbb{A}(k))^2-4\mathfrak{Det}(\mathbb{A}(k))<0$ \cite{strogatz}. However, the frequency of these oscillations, $|\lambda |\lesssim 10^{11}\,\mathrm{s}^{-1}$, is comparable to the reciprocal resolidification time and should be hardly detectable in experiments. 

The slight discrepancy in the dispersion relation cut-off between the analytical forecast (see Fig.~\ref{FigEW} and Eq.(\ref{EQ_UnstableAnalyticalCond})) and the current simulations (see Fig.~\ref{GrowthRates}) most probably arises from the 
assumption that the in-depth profiles of both the electron and lattice temperature behave as $\exp (-\xi z)$ and with the same $\xi$. Nevertheless, considering more complex profiles adds serious difficulties to the stability analysis.

\subsection{Dependencies on the parameters}

The of the temperature perturbations presented in this study is found to be directly correlated with the characteristic gradient length of the in-depth temperature profile, $\xi^{-1}$, which introduces the cut-off in the dispersion relation for high modes~\eqref{EQ_UnstableAnalyticalCond}. Two consequences can be drawn from this dependence. 

\begin{figure}
\begin{center}
\includegraphics[width=0.23\textwidth]{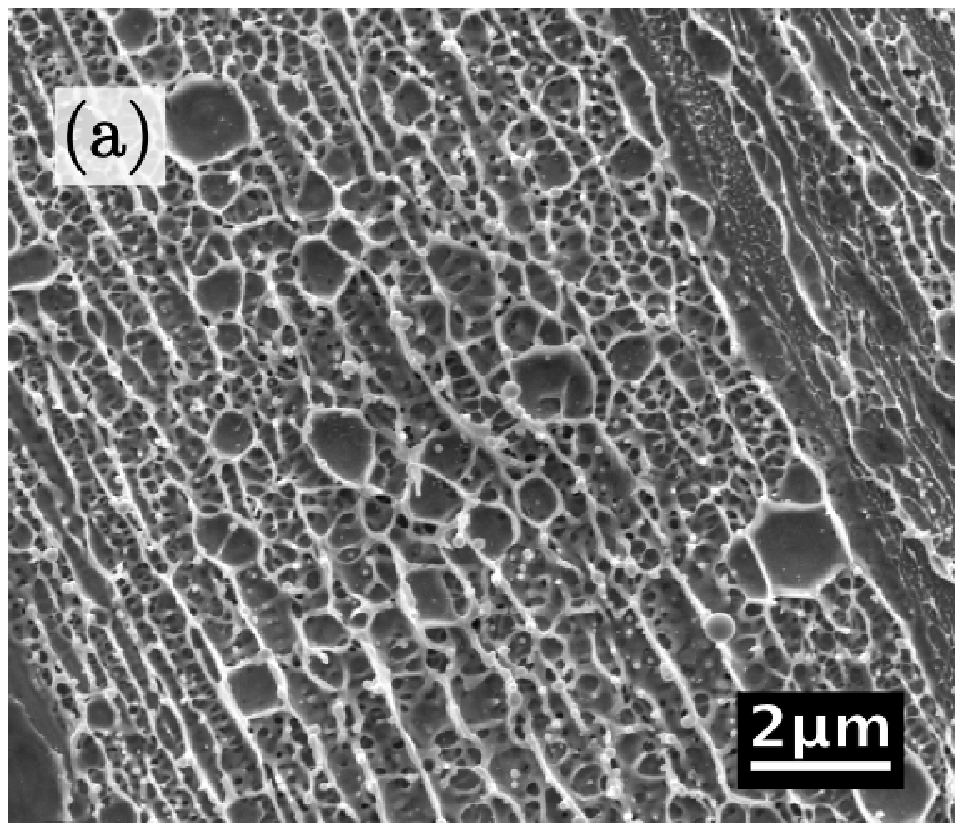}~
\includegraphics[width=0.23\textwidth]{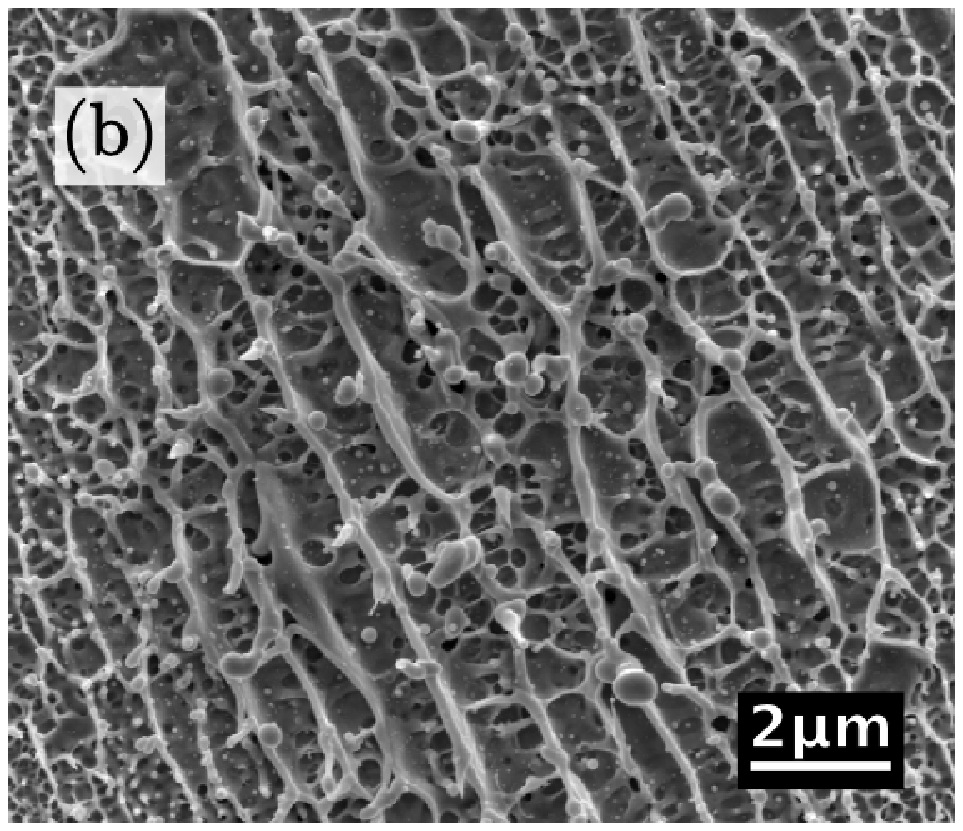}~
\includegraphics[width=0.23\textwidth]{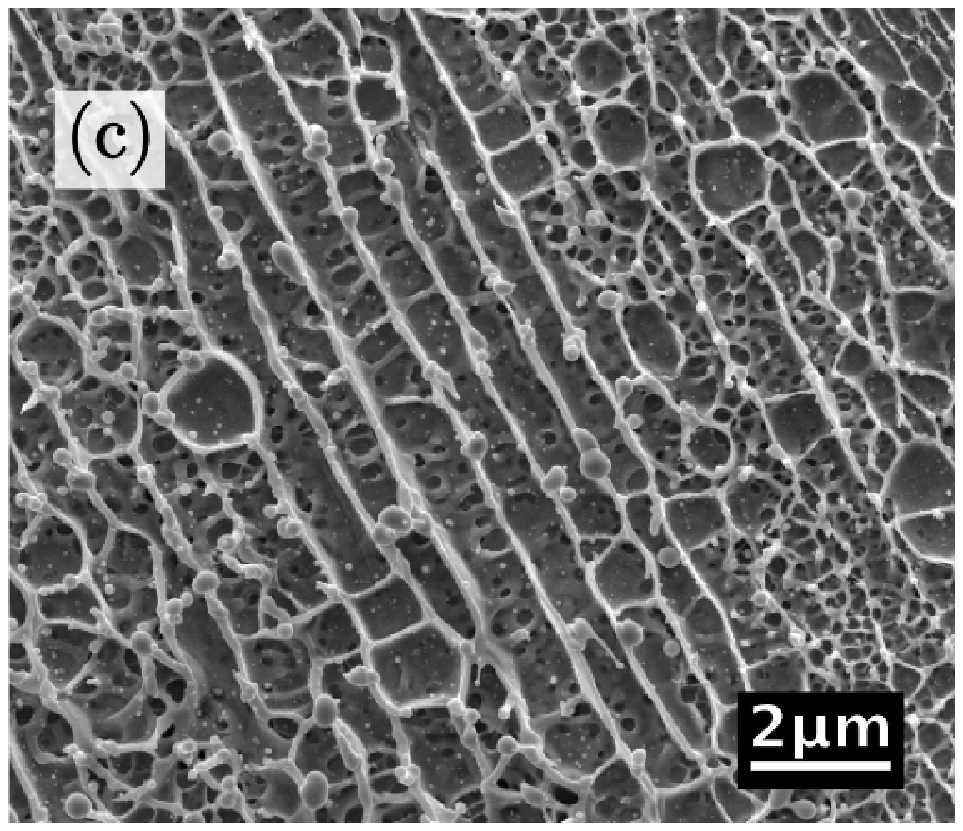}~
\includegraphics[width=0.23\textwidth]{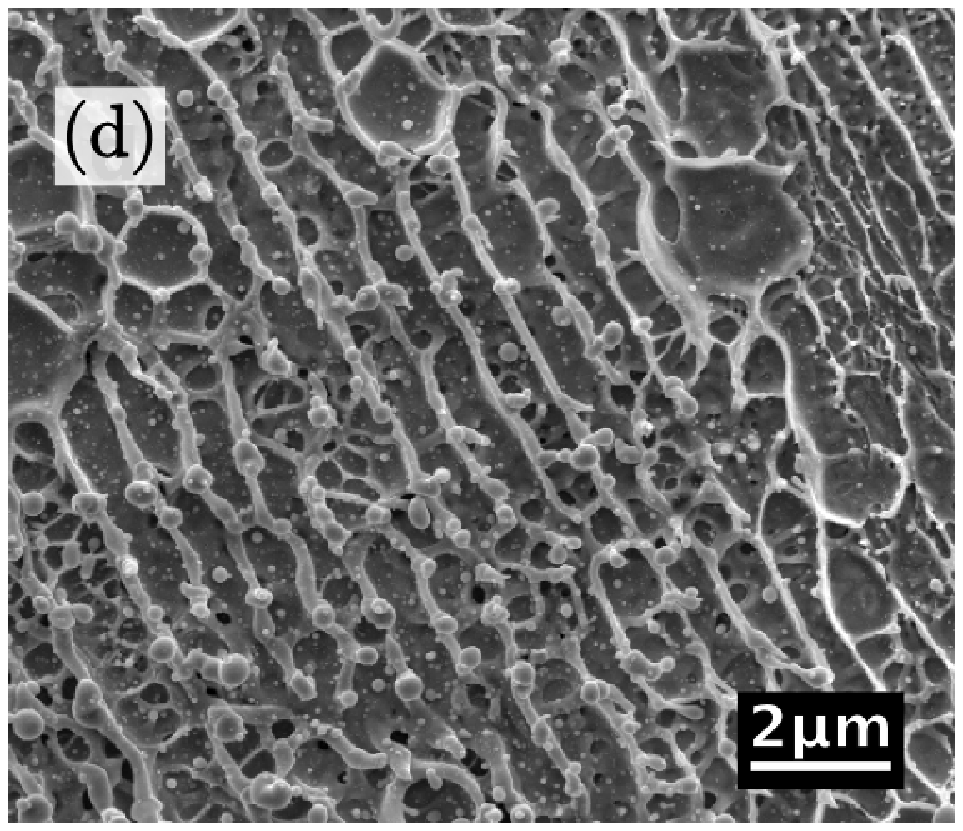}
\end{center}
\caption{Evolution of the surface pattern, which emerges on gold surfaces upon single laser pulse irradiation (Ti:sapphire, $\lambda_{\textrm{las}}=800\,$nm, pulse duration $\tau_p\approx100\,$fs, normal incidence) at different fluence values. (a) $F_0\approx 2\,\mathrm{J\cdot cm^{-2}}$, (b) $F_0\approx 3\,\mathrm{J\cdot cm^{-2}}$, (c) $F_0\approx 4\,\mathrm{J\cdot cm^{-2}}$, (d) $F_0\approx 5\,\mathrm{J\cdot cm^{-2}}$. The images correspond to the center of the laser spot; the laser beam profile was close to Gaussian.}\label{FIG_DifE}
\end{figure}

First, the $\xi$ value depends on the absorbed laser fluence $F_0$. This dependence, $\xi\propto F_0^{-1/5}$, $T_e\propto F_0^{1/2}$, was analysed by Corkum at al. \cite{corkum} for metals in a simplified case of negligible lattice thermal conductivity and large thermalization times. Hence, an increase in laser fluence will reduce the second (negative) term in the criterion~\eqref{EQ_UnstableAnalyticalCond}, whereas the first one will be less affected. This will slightly broaden the unstable band and stimulate the growth of the instability. On the other hand, if the fluence is increased, resolidification of the surface starts later and the LIPSS have more time to disappear through, e.g., surface tension.
This behaviour is confirmed in experiments. In Fig.~\ref{FIG_DifE}, the central parts of the laser-induced craters on the surface of gold are shown for different fluences of single laser shots. At low pulse energies (see Fig.~\ref{FIG_DifE}~(a)) cell-like structures dominate, while with increasing fluence the LIPSS become more and more visible (see Fig.~\ref{FIG_DifE}~(b) and~(c)). However, when the fluence becomes too large (see Fig.~\ref{FIG_DifE}~(d)), the droplets become to form on the ridges of the LIPSS pattern, making it less visible.

\begin{figure}
\begin{center}
\includegraphics[scale=0.97]{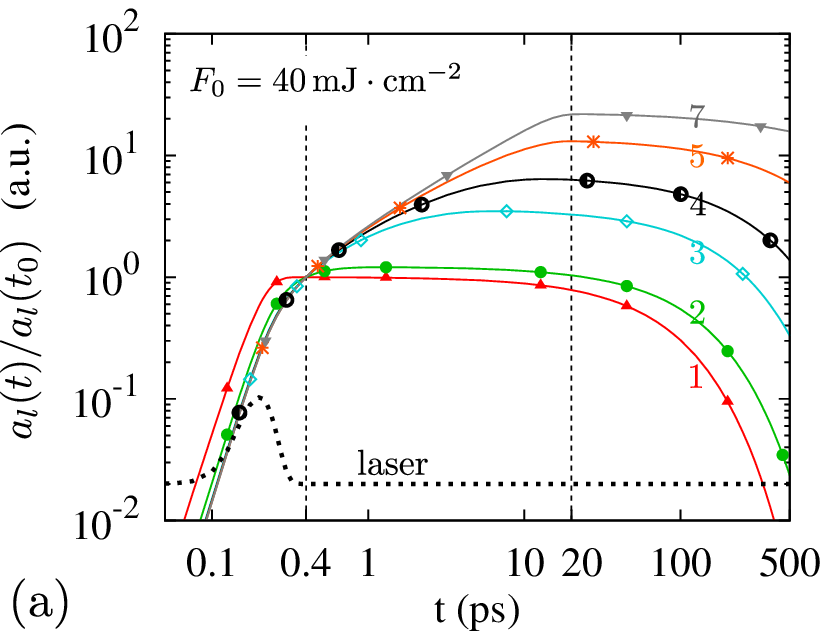} \hfill \includegraphics[scale=0.97]{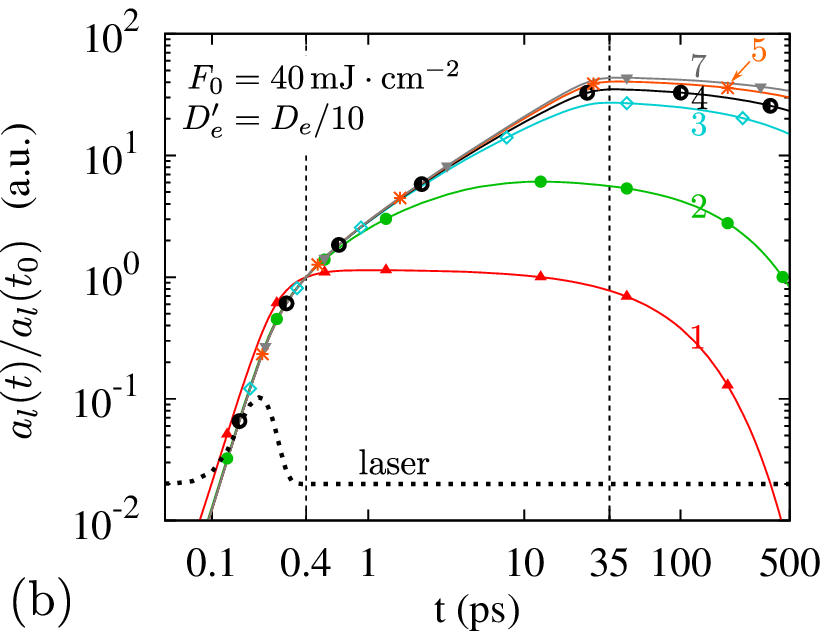}
\end{center}
\caption{Temporal evolution of the normalized amplitude of the modulations of the lattice temperature on the surface for different modulation periods (numbers correspond to Table~\ref{TAB_ModeNumbering}). (a)~Thermophysical parameters from Table ~\ref{TAB_GoldOpticThermalCoef}. (b)~Ten times smaller thermal diffusivity. The laser intensity (black dotted line) is shown in arbitrary units.}\label{FIG_Ampli_t_LowerDiffusivity}
\end{figure}

The second, more obvious, consequence resulting from the $\xi$ dependence of $k_c$ is that less diffusive materials present broader unstable spectrum. 
This feature is also confirmed by our simulations. Concentrating on gold in this paper, we tested the dependence on $\xi$ by simply changing the value of the electron thermal conductivity. The temperature gradient length, $\xi^{-1}$, is then modified together with the electron temperature diffusivity $D_e$. This artificial parametrization presents the advantage of keeping the other thermophysical quantities unchanged for easier comparison. Figure~\ref{FIG_Ampli_t_LowerDiffusivity} compares the evolution of the amplitudes of different modes obtained in the simulations performed with the parameters summarized in the Table~\ref{TAB_GoldOpticThermalCoef} and simulations performed with a new thermal conductivity $\kappa_e'$ (and therefore diffusivity, $D_e'$), which is 10 times smaller by setting $\varkappa'=31.8\,\mathrm{W/(m\cdot K)}$. 
Because of the smaller diffusivity, the region of hot electrons is more confined before the thermalization with the lattice, which consequently experiences higher heating. We therefore decreased the incoming laser fluence down to $F_0=40\,\mathrm{mJ\cdot cm^{-2}}$ to avoid material melting and facilitate the comparison. Electron-lattice themalization occurs in this case at $t_c\sim20$~ps for the parameters of Table~\ref{TAB_GoldOpticThermalCoef} and at $t_c\sim35$~ps for the case of ten times lower temperature diffusivity.

When using $\kappa_e'$, a shift of the cut-off towards high wave numbers is noticeable in Fig.~\ref{FIG_Ampli_t_LowerDiffusivity}. The mode number 2, $k=20.9\,\mathrm{\mu m^{-1}}$, is now also unstable and higher periods experience faster growth.

In these two series of simulations, the gradient length $\xi^{-1}$ of the in-depth electron temperature profile is also increasing with time between $t_0$ and $t_c$. In the simulations with the parameters from Table~\ref{TAB_GoldOpticThermalCoef}, the evolution of the $\xi^{-1}$ value at $F_0 = 40\,\mathrm{mJ\cdot cm^{-2}}$ is almost the same as at $F_0 = 110\,\mathrm{mJ\cdot cm^{-2}}$ and, in particular, $\xi^{-1}(t_0) \approx 0.15\,\mathrm{\mu m}$. The temperatures at the time moment $t_0$ are now $T_e\approx 1.1~10^4\,\mathrm{K}$ and $T_l\approx 330\,\mathrm{K}$. This provides about the same value of the criterion~\eqref{EQ_UnstableAnalyticalCond} for the normal temperature diffusivity. Regarding the simulations with $\kappa_e'$, the gradient length is in the range from $0.05\,\mathrm{\mu m}$ to $0.25\,\mathrm{\mu m}$ and, immediately after the end of the laser pulse, $T_e\approx 1.8~10^4\,\mathrm{K}$ and $T_l\approx 340\,\mathrm{K}$. This yields $k_c\approx 28.3\,\mathrm{\mu m^{-1}}$ that is in agreement with the behaviour observed in Fig.~\ref{FIG_Ampli_t_LowerDiffusivity}(b).

\section{4. Discussion}

In this paper, we have studied stability of the two-temperature model with $T_e$-dependent physical properties of material. It is demonstrated that a small initial periodic modulation of the surface temperature can start to grow if the system is driven out of thermal equilibrium. A simple analytical criterion have been obtained for the critical wave number of the instability $k_c$ as a function of the material parameters and the electron and lattice temperatures. To derive this criterion, the following simplifications were introduced:
\begin{enumerate}
\item To simplify the analytical solution, the exponential decay of the in-depth temperature profile was applied, $T\propto \exp({-\xi z})$ for both the electron and lattice temperatures, instead of $T\propto \exp({-\xi' z^2})$, as predicted by the theory \cite{corkum}; 
\item The same gradient length $\xi^{-1}$ was assumed for the electron and the lattice temperatures. In real situations, it can be larger for the electrons and, as mentioned above, is changing in time; 
\item The system was linearized, i.e., only the terms proportional to the perturbation amplitude are kept while higher-order terms are neglected in the system~\eqref{sysTsurf}.
\end{enumerate}

\begin{figure}
\begin{center}
\includegraphics[scale=0.98]{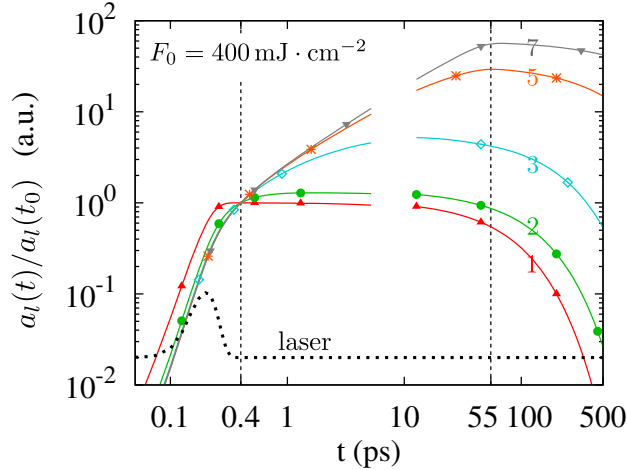}
\end{center}
\caption{Temporal evolution of the normalized amplitude of the modulation in the lattice temperature for different modes at a higher peak fluence $F_0=400\units{mJ\cdot cm^{-2}}$. The normalization has been performed with respect to the amplitude value just after the end of the laser pulse, $t_0$. Electron-lattice thermalization takes $\approx 55$ ps at this fluence.}\label{FIG_Amplitude_t_OtherFluences}
\end{figure}

Hence, the analytical estimate for $k_c$ is not exactly equal to the value calculated numerically based on the full TTM. For example for $F_0=110\,\mathrm{mJ\cdot cm^{-2}}$, the numerically calculated $k_c$ value averaged over the period $[t_0,t_c]$ is $~20\,\mathrm{\mu m}^{-1}$ (Fig.~\ref{GrowthRates}). With the gradient length of $\xi^{-1}=0.15\,\mathrm{\mu m}$ calculated by the end of the laser pulse $t_0$, Eq.~\eqref{EQ_UnstableAnalyticalCond} gives $k_c\approx 9\,\mathrm{\mu m}^{-1}$ (Fig.~\ref{GrowthRates}). At a later moment, $t\sim 2$\,ps, $\xi^{-1}\approx 0.4\,\mathrm{\mu m}$ and Eq.~\eqref{EQ_UnstableAnalyticalCond} yields $k_c\lesssim 3\,\mathrm{\mu m}^{-1}$. This should imply suppression of the growth for modes with spatial periodicity of ~2 $\mu$m and smaller (Table~\ref{TAB_ModeNumbering}) while the modes 3--5 are still growing at this time (Fig.~\ref{FIG_AmpliTlModul}). However, such a difference does not influence the main conclusions of this work and, qualitatively, the analytical criterion fits the numerical results in several aspects:
\begin{enumerate}
\item There is a clear evidence of existence of the unstable band of wave numbers;
\item The criterion~\eqref{EQ_UnstableAnalyticalCond} predicts that the width of the unstable band of the wave numbers decreases with time due to the decrease in the difference between the electron and lattice temperatures. Consequently, between the pulse termination and the coupling time, the higher modes are declining earlier than smaller ones. This is also observed numerically (see Figs.~\ref{FIG_AmpliTlModul},~\ref{FIG_Ampli_t_LowerDiffusivity} and~\ref{FIG_Amplitude_t_OtherFluences});
\item The dependence of the growth rate on the wave number calculated numerically for the full TTM (Fig.~\ref{GrowthRates}) reproduces the dependence of the maximal real eigenvalue of the system~\eqref{sysTsurf}, see Fig.~\ref{FigEW};
\item The growth rates of the instability are of the order of $10^{-11}\,\mathrm{s}^{-1}$ in both the full TTM (Fig.~\ref{GrowthRates}) and the simplified analytical (Fig.~\ref{FigEW}) models;
\item Increase in the laser fluence facilitates the instability, compare Figs.~\ref{FIG_AmpliTlModul} and~\ref{FIG_Ampli_t_LowerDiffusivity};
\item Decrease in the thermal diffusivity of the metal increases the band of unstable wave numbers, see Fig.~\ref{FIG_Ampli_t_LowerDiffusivity}.
\end{enumerate}

The explored instability-like behaviour is typical for systems driven out of thermal equilibrium~\cite{Cross}. Such behavior should manifest itself upon laser irradiation of metals when pulse duration is small enough to enable non-equilibrium between the lattice and electron temperatures on the time scale succeeding to the laser pulse. For some materials with efficient electron-lattice coupling, fast electron-lattice thermalization can significantly shorten the growth phase of the modulation amplitude of the lattice temperature. 

This periodically modulated temperature profile on the surface is a good candidate to participate in the LIPSS formation process. In view of a wide spectrum of growing modes demonstrated above, the question arises on which of them can be efficiently excited. As the LIPSS are intimately correlated with SPP excitation~\cite{Sipe, Bonse2009, Garrelie2011, derrien_rippled_2013_OE}, the SPP mechanism is very likely to constitute the first selection of the excitable modes. Even a small temperature modulation caused by a weak plasmonic wave can be amplified by the instability. Besides, the initial temperature perturbation can be induced by the surface roughness \cite{Sipe}. 
In this case, the spectrum of the initial perturbations can be broad. The proposed instability then serves as a low-pass filter with the limiting frequency described by the equation~\eqref{EQ_UnstableAnalyticalCond}.
This should enhance a part of the initial spectrum with $k<k_c$ but smoothen the high-frequency part.

A prerequisite for the instability-assisted scenario of the LIPSS formation is that the modulations amplitude of the lattice temperature can remain significant up to the temperature levels when ablation or, at least, melting occurs, which can imprint the surface relief on the sample either via periodic ablation or molten material relocation. Figures~\ref{FIG_AmpliTlModul}, \ref{FIG_Ampli_t_LowerDiffusivity} (a), and~\ref{FIG_Amplitude_t_OtherFluences} demonstrate that nearly the same unstable spectrum for gold exists at different fluences, even above the transition to melting. In the Fig.~\ref{FIG_Amplitude_t_OtherFluences}, the interruption in the curves of the lattice temperature amplitude (normalized to $a_l(t_0)$) are the signature of a partially molten material on the surface. The melting process is associated with a stagnation of the lattice at the melting temperature until the enthalpy of fusion is locally provided (see Ref.~\cite{levy_relaxation_2016_ASS} for details), after which the temperature can further grow. Within the molten phase at the melting point, the modulation amplitude is thus 0 which explains the discontinuity in the curves of Fig.~\ref{FIG_Amplitude_t_OtherFluences} in log scale. The modulation of the absorbed laser energy is hidden in the form of the spatially modulated molten fraction of material. When the melting process has been completed and material can be further heated, the temperature modulation reappear. It should be mentioned that, if the molten layer survives sufficiently long on the sample surface, relocations of liquid material are likely to deform the original modulation and modify the amplitudes. However, the relocation will simultaneously lead to the formation of the surface relief. The detailed analysis of this hydrodynamic process is beyond the scope of this paper \cite{gurevich_hydro_2016_ASS}.

It is necessary to underline that the temperature-dependent reflection coefficient can provides an additional positive-feedback mechanism, which facilitates the instability of the surface temperature profile.
All above-presented simulations were performed for absorbed fluence. This implicitly assumes a constant reflectivity of the irradiated sample. However, due to strong non-equilibrium excitation of the electron subsystem at ultrashort laser pulses, the reflectivity of metals can considerably be changing during irradiation. When the electron temperature increases but still remains under the Fermi temperature (about 5.5~eV for gold~\cite{AshcroftMermin}), the characteristic collision frequency of electrons is also increasing that is associated with an enhanced light absorption. Consequently, the regions where the peaks of the temperature modulations develop (primarily for the electron temperature) become more absorbing during the laser pulse that facilitates their heating. In the regions of lower electron temperatures (in the modulation troughs), the reflectivity also decreases but to a smaller extent than in the modulation peaks. As a result, a positive feedback mechanism develops during the laser pulse which can be expected to enhance the growth rate of the perturbations.

This has been neglected in the analytical modelling to avoid severe mathematical complications. However, the variations of optical properties can be described in the frames of the Drude model in our numerical treatment.

To study the role of optical feedback effect, two sets of simulations have been performed. The first one considers the constant reflectivity, $R=0.974$~\cite{Palik1985}, of a flat, ideally polished gold sample irradiated in vacuum by 100-fs, 800-nm laser pulse at normal incidence. The second set of modelling includes the Drude model for optical properties of gold to calculate the reflection and absorption coefficients as a function of the lattice and electron temperatures~\cite{kirkwood_experimental_2009_PRB}. The real and imaginary parts of the dielectric function can respectively be expressed as $\varepsilon' = 1 - \omega_{pe}^2 / \left( \omega^2 + \nu_\textrm{eff}^2 \right)$ and $\varepsilon'' = \left( \nu_\textrm{eff}\omega_{pe}^2 \right)/ \left(\omega \left(\omega^2 + \nu_\textrm{eff}^2 \right) \right)$ with $\omega_{pe}$ the plasma frequency, $\omega = 2\pi c/\lambda_{\textrm{las}}$, and $\nu_\textrm{eff} = \min \left( \nu_e ; \nu_c\right)$. The electronic collision frequency is the sum of the electron-electron and electron-phonon collision frequencies~\cite{Lin}, $\nu_e = AT_e^2 + B T_l$. $\nu_c$ is the maximal, critical collision frequency estimated by the inter-atomic distance in the material of density $n_0$ and the electron mean velocity depending on the Fermi velocity $v_F$ as $\nu_c=n_0^{1/3}\sqrt{v_F^2+\frac{k_BT_e}{2m_e}}$. The parameters $A=1.2\times 10^{7}\,\mathrm{K^{-2}s^{-1}}$ and $B=1.23\times 10^{11}\,\mathrm{K^{-1}s^{-1}}$ are taken from Ref.~\cite{wang_timeresolved_1994_PRB}. 
From the above relations, it is possible to calculate $\varepsilon '$ and  $\varepsilon ''$ as a function of the temperatures provided that the plasma frequency is known.
In particular, at room temperature, $T_r$, at thermal equilibrium,  $\varepsilon'(T_e,T_l)$ and  $\varepsilon''(T_e,T_l)$ must be equal to the experimentally measured values \cite{Palik1985}, $\varepsilon'=-26.2$ and $\varepsilon''=1.85$. This condition is used to estimate the missing parameter 
$\omega_{pe}=\omega\sqrt{1-\varepsilon'(T_r) +\varepsilon''(T_r)^{\,2}/ \left(1-\varepsilon'(T_r) \right)}$. From the local estimation of $\varepsilon(T_e,T_l)$ in the simulations, the absorption and reflection coefficients, $\alpha (x,z,t)$ and $R(x,t)=\left|(1-\sqrt{\varepsilon})/(1+\sqrt{\varepsilon})\right|^2$ respectively, can be calculated that self-consistently modifies the locally absorbed fluence.

\begin{figure}
\begin{center}
\includegraphics[scale=0.97]{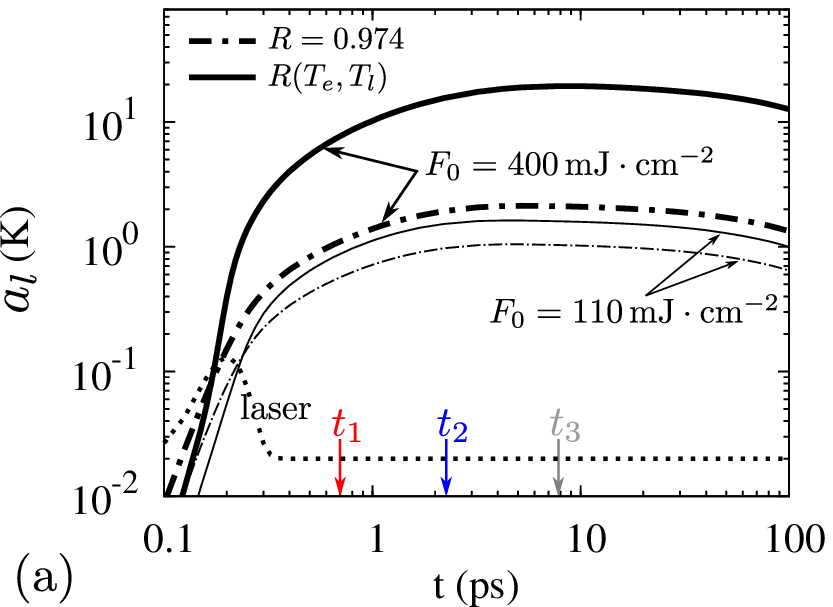}\hfill
\includegraphics[scale=0.97]{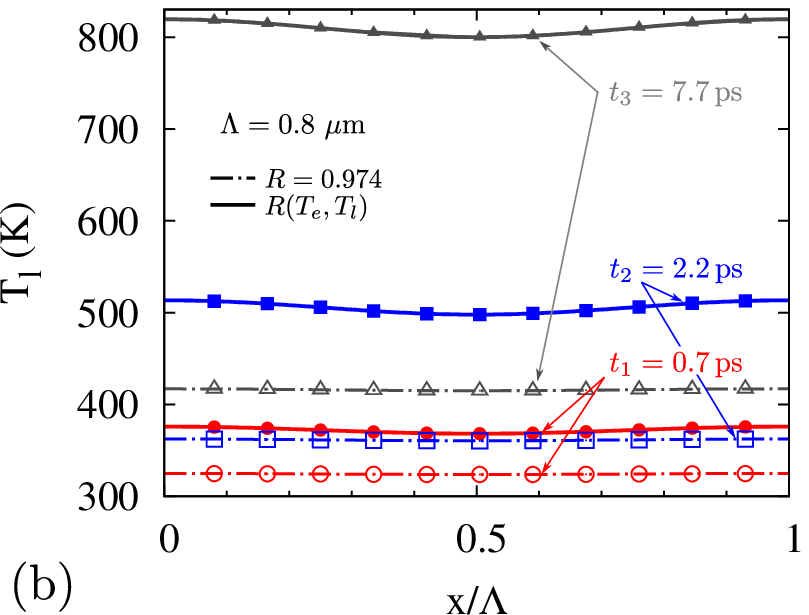}
\end{center}
\caption{Comparison of the modulation dynamics of the lattice temperature for $\Lambda=\Lambda_\textrm{exp}=0.8\,\mathrm{\mu m}$ calculated with the constant (dot-dashed lines) and temperature-dependent (solid lines) optical properties. The initial laser intensity amplitude is $\eta = 0.05$. (a) Time evolution of the modulation amplitude,  $a_l$, on the surface of gold at $F_0 = 110\,\mathrm{mJ\cdot cm^{-2}}$ (thin lines) and $F_0 = 400\,\mathrm{mJ\cdot cm^{-2}}$ (thick lines). (b) Lattice temperature profiles along the surface at different time moments for~$400\,\mathrm{mJ\cdot cm^{-2}}$.}\label{FIG_Amplitude_ChangeInOpticalProp}
\end{figure}

Figure~\ref{FIG_Amplitude_ChangeInOpticalProp} shows the evolution of the modulation amplitude of the lattice temperature developed from the perturbation initially introduced into laser intensity (Eq. \eqref{EQ_TransverseModulations}) with the spatial period of $0.8\,\mathrm{\mu m}$ and the relative amplitude $\eta = 0.05$. It is clear that the amplitude of the unstable mode is considerably emphasized when accounting for the decrease in the optical properties during the laser pulse irradiation, leading to enhancement of the modulation amplitude up to an order of magnitude in the case of~$400\,\mathrm{mJ\cdot cm^{-2}}$.


The transient optical response of metals can thus facilitate the growth of the amplitude of the unstable modes. Its impact becomes more distinct at high laser fluences which makes it an important enhancement process in the regime of single-pulse LIPSS formation as demonstrated in Figs.~\ref{LipssAu} and~\ref{FIG_DifE}. 

Finally, we mention that the periodic modulation of the lattice temperature can induce an additional impact on the LIPSS formation process through generation of modulated thermal stress. The modulated stress should favor material relocation from the temperature peaks to the colder regions, assisting in the development of the periodic surface relief. The studies of the possible role of modulated thermal stress in the LIPSS formation is under progress.


\section{5. Conclusions}

In this paper, we demonstrate analytically and numerically that the two-temperature model can be unstable with respect to a small initial temperature modulation on a metal surface at the regimes of ultrafast laser excitation. 
The characteristic time of growth of the modulated temperature pattern, $1/\gamma$, is of the order of 10~picoseconds; hence an inhomogeneous temperature distribution along the surface is already developed by the time when melting and ablation take place. The growth of the surface temperature modulation gives a convincing explanation of the periodically-modulated surface profile  (LIPSS) upon  single-pulse laser ablation, see Fig.~\ref{LipssAu}. This model does not contradict but extends the existing models of the LIPSS formation (interference of incident and surface-scattered electromagnetic waves, surface plasmon polaritons). These models provide physical reasons for the initial modulation of the temperature on the surface. From this viewpoint, the formation of the LIPSS pattern can be split in three stages: (1) Formation of an initial modulation of the surface temperature, e.g., due to interaction of the incident light with the SPP or with the wave scattered on the surface roughness. We stress that the amplitude of this initial modulation can be relatively small at this stage. (2) Amplification of the temperature modulation by the instability described in this paper. (3) Relocation of the melt induced by the high-amplitude temperature modulation whose mechanisms should be explained and described in the frames of hydrodynamics and instabilities developed in thin films of molten material.

\section*{ACKNOWLEDGMENTS}

This work was partially supported by several funding agencies. NMB and YL acknowledge the support from the
state budget of the Czech Republic (project HiLASE: Superlasers for the real world: LO1602).
SVG acknowledges the support of the Center for Nonlinear Science (CeNoS) of the University of M\"unster.

%

\end{document}